\documentclass[usenatbib,usegraphicx,onecolumn]{mn2e}

\def\apj{ApJ}
\def\apjl{ApJL}

\def\aap{A\&A}
\def\aaps{A\&AS}

\def\mnras{MNRAS}
\def\aj{AJ}
\def\nat{Nature}
\def\araa{ARA\&A}
\def\prd{Ph. Rv. D}
\def\apss{ApSS}
\def\cm{\textrm{cm}}

\def\pc{\textrm{pc}}
\def\Mpc{\textrm{Mpc}}

\def\Jy{\textrm{Jy}}

\def\kms{\textrm{km}~\textrm{s}^{-1}}

\def\gcm2{\textrm{g}~\textrm{cm}^{-2}}

\def\phcm2s1{\textrm{photons}~\textrm{cm}^{-2}~\textrm{s}^{-1}}

\def\eV{\textrm{eV}}

\def\MeV{\textrm{MeV}}
\def\GeV{\textrm{GeV}}

\def\MHz{\textrm{MHz}}
\def\GHz{\textrm{GHz}}

\def\Myr{\textrm{Myr}}

\def\kyr{\textrm{kyr}}
\def\muGauss{\mu\textrm{G}}
\def\mGauss{\textrm{mG}}

\newcommand{\mean}[1]{\ensuremath{\langle #1 \rangle}}

\title[Equipartition Magnetic Fields for Starbursts]{The Equipartition Magnetic Field Formula in Starburst Galaxies: Accounting for Pionic Secondaries and Strong Energy Losses}
\author[Lacki \& Beck]{Brian C. Lacki$^{1,2}$ and Rainer Beck$^3$\\$^1$Jansky Fellow of the National Radio Astronomy Observatory\\$^2$Institute for Advanced Study, Einstein Drive, Princeton, NJ 08540, USA, brianlacki@ias.edu\\$^3$Max-Planck-Institut f\"ur Radioastronomie, Auf dem H\"ugel 69, 53121 Bonn, Germany}

\date{Draft Version}

\begin{document}

\maketitle

\begin{abstract}
Equipartition arguments provide an easy way to find a characteristic scale for the magnetic field from radio emission, by assuming the energy densities in cosmic rays and magnetic fields are the same.  Yet most of the cosmic ray content in star-forming galaxies is in protons, which are invisible in radio emission.  Therefore, the argument needs assumptions about the proton spectrum, typically that of a constant proton/electron ratio.  In some environments, particularly starburst galaxies, the reasoning behind these assumptions does not necessarily hold: secondary pionic positrons and electrons may be responsible for most of the radio emission, and strong energy losses can alter the proton/electron ratio.  We derive an equipartition expression that should work in a hadronic loss-dominated environment like starburst galaxies.  Surprisingly, despite the radically different assumptions from the classical equipartition formula, numerically the results for starburst magnetic fields are similar.  We explain this fortuitous coincidence using the energetics of secondary production and energy loss times.  We show that these processes cause the proton/electron ratio to be $\sim 100$ for GHz-emitting electrons in starbursts.
\end{abstract}


\section{Introduction}
\label{sec:Introduction} Magnetic fields are important in astrophysics, but in practice their strengths are very hard to directly determine.  Their presence is inferred through synchrotron emission of cosmic ray (CR) electrons and positrons ($e^{\pm}$) gyrating in magnetic fields.  Synchrotron emission is detected from star-forming galaxies, demonstrating that they have magnetic fields \citep[e.g.,][]{Condon92,Beck05-Review}.  However, the synchrotron emission only informs us of a combination of cosmic ray content and the magnetic field content.  In a few cases for star-forming galaxies, a detected gamma-ray spectrum provides additional evidence of the CR energy density (\citealt{Aharonian06}; \citealt{Abdo10-M31} and references therein), which when combined with plausible assumptions and modelling, allows us to constrain the magnetic fields in galaxies \citep[e.g.,][]{deCeaDelPozo09-Obs,Crocker10-BLimit,Crocker11-Wild,Lacki10-XRay}.  The great majority of galaxies are undetected in gamma rays, though (some upper limits are given in \citealt{Lenain11} and \citealt{Ackermann12}).  Faraday measurements, which require a polarization signal, are a useful way of measuring ordered magnetic field strengths.  Unfortunately, starbursts likely have highly turbulent magnetic fields, so the expected polarization is small.  Anisotropy introduced in the turbulence by shearing and compression \citep{Laing80,Sokoloff98,Beck12} may result in some polarization signal (c.f. \citealt{Greaves00,Jones00}), but there still will be no Faraday signal \citep[e.g.,][]{Beck05-Review}.  Furthermore in starbursts, Faraday depolarization can remove any polarization signal at typical observing frequencies \citep{Reuter94,Sokoloff98}.  Faraday rotation measures also can be biased if there are relationships between the magnetic field strength and density, or if anisotropic turbulence is present on the sightline \citep{Beck03}.

The equipartition and minimum energy arguments are common ways of finding a characteristic magnetic field strength $B$ \citep[e.g.,][]{Beck01}.  For a given radio luminosity, if $B$ is very small, then the CR energy density $U_{\rm CR}$ must be very large; likewise, if $U_{\rm CR}$ is very small, then $B$ must be very high.  In between, there is a single magnetic field strength where $U_B = B^2 / (8 \pi)$ is equal to $U_{\rm CR}$: this is the equipartition magnetic field strength ($B_{\rm eq}$).  Alternatively, for a given radio flux, the combined nonthermal energy density $U_B + U_{\rm CR}$ has a minimum at a magnetic field strength $B_{\rm min}$ which is of the same order as (though generally distinct from\footnote{When neither CR protons nor CR $e^{\pm}$ are cooled, the minimum energy and equipartition magnetic field strengths are equal if CRs are injected with a spectral index of 3 \citep{Beck05}.}) $B_{\rm eq}$ \citep{Burbidge56-Equip}.

Throughout much of the Milky Way, equipartition between the magnetic fields and CRs holds \citep[e.g.,][]{Niklas97}.  Likewise, in other normal star-forming galaxies, equipartition likely holds to within a factor of $\sim 10$ in $B$ \citep[e.g.,][]{Duric90}.  It is less clear if equipartition holds in starbursts, although any deviation would be interesting in its own right for the propagation of CRs and the sources of the magnetic fields (e.g., \citealt{Thompson06}; \citealt*{Lacki10-FRC1}).  Equipartition has been proposed as a cause of the observed correlation between the far-infrared and radio luminosities of star-forming galaxies \citep{Niklas97}.  On the other hand, \citet{Thompson06} argued that equipartition (between magnetic fields and CRs) predicts magnetic field strengths too low to allow starbursts to lie on the correlation, since other radiative losses are extremely fast and $B$ must be large for thereto be any significant synchrotron emission \citep[see also][]{Lacki10-FRC1}.  However, the ease of equipartition methods -- compared to methods involving gamma rays, detailed spectral modelling, or polarization methods -- has led to their application in a variety of environments including normal galaxies, the Galactic Center \citep[e.g.,][]{LaRosa05,Ferriere09}, starburst galaxies (e.g., \citealt*{Volk89-Equip}; \citealt{Beck05-Starburst,Persic10,Beck12}), and galaxies at high redshift \citep{Murphy09-zHigh,Chakraborti12}.

A problem with equipartition-style estimates is that most of the CR energy density is actually invisible in synchrotron emission: CR protons (and nuclei) are the dominant CR population at high energies.  Only in the few cases where gamma-ray observations are available can the proton energy density be constrained directly \citep{Acero09,Acciari09,Lacki11-Obs,Persic12}. Radio equipartition estimates therefore make assumptions about how to convert the observed CR electron population in some observed frequency range into the total CR proton energy density.  The simplest and most common assumption is to assume that the CR proton and electron spectra are power laws in energy with the same spectral index, and that there is a single ratio $\kappa = 30 - 100$ that sets the ratio for all galaxies.  This is likely to be true for the injection spectra of primary CRs \citep{Bell78,Schlickeiser02}, at least those associated with star formation.  The spectrum of Milky Way CRs indicates the primary injection proton/electron is near $\sim 100$ \citep{Ginzburg76}.

The problem with this assumption is that the CR spectrum can be complex, especially for CR $e^{\pm}$.  Roughly speaking, the steady-state CR spectrum $N(E)$ is equal to the product of the injection spectrum $Q(E)$ and the characteristic loss (cooling or escape) time $t_{\rm loss} (E)$. When CR protons and electrons are governed by the same losses, such as diffusive escape in normal galaxies, the proton/electron ratio is preserved.  In starburst galaxies, however, the CRs are likely to experience a variety of energy losses with different energy dependencies, which can alter the actual proton/electron ratio \citep{Beck05}.  At high energies, $e^{\pm}$ are cooled quickly by synchrotron and Inverse Compton (IC) losses, which increases the proton/electron ratio with energy.  This is seen in the Milky Way at energies $\ga 10\ \GeV$ (see for example Figure 3.29 of \citealt{Schlickeiser02}).  In addition, there are ionization losses for both $e^{\pm}$ and protons, bremsstrahlung losses of $e^{\pm}$, and winds, which can alter the CR $e^{\pm}$ spectrum \citep{Thompson06,Murphy09-zHigh,Lacki10-FRC1}.

A further complication is the possible presence of pionic secondary $e^{\pm}$ in starburst galaxies.  Secondary $e^{\pm}$ are generated when protons crash into ambient gas atoms, creating unstable pions that decay into gamma rays, neutrinos, and $e^{\pm}$.  In the dense gases of starbursts, the amount of secondaries may be comparable to or even dominant over the primary electrons (e.g., \citealt{Torres04-Arp220,Rengarajan05}; \citealt*{Thompson07}; \citealt{Lacki10-FRC1}).

In this paper, we present equipartition and minimum energy formulae that should work in starburst environments.  These take into account the presence of secondary $e^{\pm}$ and the strong energy losses of starburst galaxies.

\section{Derivation of Equipartition and Minimum Energy Magnetic Fields in Starbursts}
\subsection{Basic assumptions}
The spectrum $N(E)$ of CRs is governed by the diffusion-loss equation:
\begin{equation}
\frac{\partial N(E)}{\partial t} = Q(E) + \frac{d}{dE}[b(E)N(E)] - \frac{N(E)}{\tau(E)} + D \nabla^2 N(E)
\end{equation}
as given in \citet{Torres04-Arp220}.  Here $Q(E)$ is the injection
spectrum of CRs, $b(E)=-dE/dt$ is the cooling rate of individual
CRs, $\tau(E)$ is the loss time to escape and losses that
remove most of the CR's energy in one interaction (catastrophic
losses), and $D$ is the spatial diffusion constant.  We can
simplify the equation to the leaky box equation if we assume that
the modelled region is homogeneous, so that all spatial terms drop
out, and is in equilibrium ($\partial N(E)/\partial t = 0$).  The
leaky box equation is then
\begin{equation}
0 = Q(E) + \frac{d}{dE}[b(E)N(E)] - \frac{N(E)}{\tau(E)}
\end{equation}
where escape out of the region is now considered a catastrophic loss.

In the test particle approach to CR acceleration, CRs are injected
with a momentum power law spectrum \citep{Bell78}.  To simplify
matters, we assume that CRs are injected with an energy power law
spectrum:
\begin{equation}
Q(E) = Q_0 \left(\frac{E}{mc^2}\right)^{-p},
\end{equation}
where $E$ is the total (rest plus kinetic) energy of the particle.
The power law spectrum continues to a maximum energy $\gamma_{\rm
max} m c^2$.

At high energies, protons escape or lose energy through pion
production. Since pion production can be modelled as a catastrophic
loss \citep{Torres04-Arp220}, the proton spectrum is simply
\begin{equation}
\label{eqn:pSpectrumSoln}
N_p(E) = Q_p(E) \tau_p (E).
\end{equation}
The proton lifetime in starbursts $\tau_p$ is thought to be
dominated by advection or possibly pionic losses in denser
starbursts like Arp~220.  The fact that starbursts are observed at
TeV energies \citep{Acero09,Acciari09} indicates hard gamma-ray
spectra ($\Gamma \approx 2.2$), supporting the idea that $\tau_p$ is
set by one of these energy-independent processes.  The gamma-ray
observations of M~82 and NGC~253 suggest that more CR proton power
is advected away in their nuclear starburst winds than is lost to
pion production, although the efficiency of pionic losses is still
much greater than in the Milky Way \citep{Lacki11-Obs}.  As we will
see, though, it is convenient to scale $\tau_p$ with the pionic
lifetime $t_{\pi}$, because the secondary $e^{\pm}$ injection rate
is directly tied to the pionic loss time.  We therefore parameterize
the proton lifetime as:
\begin{equation}
\tau_p = F_{\rm cal} t_{\pi}.
\end{equation}

CR $e^{\pm}$ in contrast are thought to largely be trapped in
starburst galaxies \citep{Volk89-Calor}, although winds may be quick
enough to remove $e^{\pm}$ before they cool in some starbursts
\citep{Heesen11}.  Their losses are therefore continuous.  The total
$e^{\pm}$ energy loss rate $b(E)$ can also be written as a cooling
time $t_{e} (E) = E / b(E)$, giving us a steady-state spectrum
\begin{equation}
\label{eqn:eSpectrumSoln}
N_e(E) = \frac{Q_e(E) E}{(p - 1) b(E)} = \frac{Q_e(E) t_{e} (E)}{p - 1}.
\end{equation}

To get the total CR proton energy density, we integrate its kinetic
energy over all allowed energies: $U_p = \int_{m_p c^2}^{\gamma_{\rm
max} m_p c^2} N_p(E) K dE$.  Substituting in the solutions for the
steady-state and injection proton spectra, we find
\begin{equation}
U_p = t_{\pi} F_{\rm cal} Q_0 (m_p c^2)^2 \Upsilon
\end{equation}
where we define a factor
\begin{equation}
\Upsilon  = \left\{ \begin{array}{ll} \displaystyle \frac{1 - \gamma_{\rm max}^{2 - p}}{p - 2} + \frac{1 - \gamma_{\rm max}^{1 - p}}{p - 1} & (p \ne 1, 2)\\
                                      (\ln \gamma_{\rm max} + 1 - 1/\gamma_{\rm max}) & (p = 2) \end{array} \right.
\end{equation}
that depends on the shape of the injection spectrum. Unless
otherwise noted in the paper, we assume $\gamma_{\rm max} = \infty$
and parameterize $p$ as $2.2 + \Delta p$, so that $\Upsilon =
4.1\bar{6} = 25 / 6$ when $\Delta p = 0$.  We can compare this to
the luminosity of protons at energy $E$, which can be approximated
as $E^2 Q_p(E)$.  We find
\begin{equation}
\label{eqn:UpGivenE2Qp}
U_p = \left(\frac{E_p}{m_p c^2}\right)^{p - 2} \Upsilon t_{\pi} F_{\rm cal} \times E_p^2 Q_p(E)
\end{equation}

\subsection{Relating the secondary $e^{\pm}$ to CR protons}
\label{sec:ProtonsToSecondaries} Calculating the secondary $e^{\pm}$
spectrum involves integrating the differential cross sections for
pionic $e^{\pm}$ production over the CR proton spectrum (e.g.,
\citealt{Kamae06}; \citealt*{Kelner06}).  However, at sufficiently
high $e^{\pm}$ energies, the secondary $e^{\pm}$ spectrum can be
estimated using a $\delta$-function approximation.  A fraction
$F_{\rm cal}$ of the CR proton energy goes into pion production,
with the rest either escaping or lost to ionization or Coulomb
cooling.  Charged pions makes up $\sim 2/3$ of the pions produced,
with the remaining $1/3$ going into neutral pions that mainly decay
into gamma rays.  Assuming equal energy going into the pion decay
products, approximately $1/4 \times 2/3 \approx 1/6$ of the power in
pions ends up in secondary $e^{\pm}$
\citep[c.f.,][]{Burbidge56-Secondaries,Loeb06}.  The other half of
the pionic luminosity ($3/4$ of the charged pion power) goes
into neutrinos.  For the $\delta$-function approximation, we
therefore have:
\begin{equation}
\label{eqn:SecondarySpectrum}
E_{\rm sec}^2 Q_{\rm sec} (E_{\rm sec}) = \frac{F_{\rm cal}}{6} E_p^2 Q_p (E_p)
\end{equation}
Since the typical inelasticity per inelastic collision is 20 per
cent, and each pion decays into four particles, the typical energy
of a secondary $e^{\pm}$ is roughly $E_{\rm sec} \approx 0.05 E_p$.

Do secondaries in fact dominate the $e^{\pm}$ spectra of starbursts?
With equation~\ref{eqn:SecondarySpectrum}, we can estimate the
primary to secondary ratio:
\begin{equation}
\frac{Q_{\rm sec} (E_e)}{Q_{\rm prim} (E_e)} \approx K_0 \frac{F_{\rm cal}}{6} \left(\frac{E_p}{E_e}\right)^{2 - p}
\end{equation}
where we define $K_0 = Q_p / Q_{\rm prim}$ as the number ratio of
injected primary protons and primary electrons at high energies ($\ga m_p
c^2$).  While the injection rate of protons vastly overwhelms that
of primary electrons at a constant energy, the secondary $e^{\pm}$
are injected at much lower energy than their parent protons, where
there is more power in primary electrons for $p > 2$.  Thus the
secondary $e^{\pm}$ are `diluted' with respect to the primary
electrons \citep[e.g.,][]{Lacki10-FRC1}.  For $p = 2.2$, we find
\begin{equation}
\frac{Q_{\rm sec} (E_e)}{Q_{\rm prim} (E_e)} \approx 6.9 F_{\rm cal} \left(\frac{K_0}{75}\right).
\end{equation}
In terms of the fraction of the total $e^{\pm}$ population in
secondaries, $f_{\rm sec} = Q_{\rm sec} / (Q_{\rm sec} + Q_{\rm
prim}) = [1 + (Q_{\rm sec} / Q_{\rm prim})^{-1}]^{-1}$, or
\begin{equation}
f_{\rm sec} \approx \left[1 + 0.15 F_{\rm cal}^{-1} \left(\frac{K_0}{75}\right)^{-1}\right]^{-1}
\end{equation}
for $p = 2.2$.

The gamma-ray spectra of M~82 and NGC~253 indicate that $F_{\rm
cal} \approx 0.2 - 0.5$ \citep{Lacki11-Obs}, so our equations imply
secondary to primary ratios of $\sim 1.4 - 3.4$, or $f_{\rm sec}
\approx 0.6 - 0.8$.  This is in line with detailed modelling of
these galaxies (\citealt{Domingo05}; \citealt*{deCeaDelPozo09-M82};
\citealt*{Rephaeli10}).

\subsection{Relating the radio emission to the $e^{\pm}$ spectrum}
We also consider a $\delta$-function approximation in calculating the
synchrotron emission.  In this approximation, we assume each
$e^{\pm}$ of energy $E_e$ radiates all of its synchrotron emission
at a frequency:
\begin{equation}
\label{eqn:nuC}
\nu = \frac{3 E_e^2 e B \mean{\sin \alpha}}{4 \pi m_e^3 c^5} = \frac{3 E_e^2 e B}{16 m_e^3 c^5},
\end{equation}
with a magnetic field strength $B$ and an electron electric charge
$e$ \citep{Rybicki79}. We have used the fact that the mean sine
pitch angle $\mean{\sin \alpha}$ for an isotropic distribution of CR
$e^{\pm}$ is $\mean{\sin \alpha} = \pi / 4$.  While physically
speaking the $\delta$-function approximation is not correct since
the synchrotron emission of an $e^{\pm}$ is a broad continuum, it
gives acceptably accurate results for a power law distribution of
$e^{\pm}$ energies \citep{Felten66}.

Since an $e^{\pm}$ would radiate its energy in a synchrotron
lifetime $t_{\rm synch}$, the volumetric synchrotron luminosity of a
homogeneous region is $\nu \epsilon_{\nu} = d\epsilon/d\ln\nu =
(dU_e/d\ln\nu) / t_{\rm synch} = (E_e dN_e/d\ln\nu) / t_{\rm
synch}$, where $\nu$ is the observed frequency.  From
equation~\ref{eqn:nuC}, $d\ln \nu = 2 d\ln E_e$, so we have $\nu
\epsilon_{\nu} \approx 1/2 (E_e dN/d\ln E_e) / t_{\rm synch} = 1/2
(E_e^2 N_e (E_e)) / t_{\rm synch}$.  The steady-state spectrum of
the electrons is given by equation~\ref{eqn:eSpectrumSoln};
substituting that in, we get
\begin{equation}
\label{eqn:nuLnuFromE2Qe}
\nu \epsilon_{\nu} = \frac{1}{2} E_e^2 Q_e (E_e) \frac{g t_{e} (E_e)}{(p - 1) t_{\rm synch} (E_e)}.
\end{equation}
The factor $g$ corrects our approximation for $\nu
\epsilon_{\nu}$ to the exact result for a power law spectrum of
ultrarelativistic $e^{\pm}$ at the frequency $\nu_C$
\citep{Rybicki79}.  If the electron lifetime is dominated by a
continuous energy loss process and scales as $t_e (E_e) \propto
E_e^a$, then the value of $g$ is:
\begin{equation}
g = \frac{27 \sqrt{3}}{(p - a + 1)} 2^{-3(5 + a - p) / 2} \pi^{(2 + a - p)/2} \Gamma\left(\frac{5 + p - a}{4}\right) \Gamma\left(\frac{3p - 3a - 1}{12}\right) \Gamma\left(\frac{19 + 3p - 3a}{4}\right) \left[\Gamma\left(\frac{7 + p - a}{4}\right)\right]^{-1}.
\end{equation}
We compile the values of $g$ for different loss processes and
injection indices $p$ in Table~\ref{table:gValues}, but for
bremsstrahlung, synchrotron, or IC cooling, $g$ is very nearly 1,
and it is exactly 1 when $p = 2.0$ and $a = -1.0$ (as in the case
when synchrotron cooling sets the $e^{\pm}$ lifetime).

\begin{table*}
\begin{minipage}{170mm}
\caption{Values of $g$ for synchrotron emission spectra}
\label{table:gValues}
\begin{tabular}{lcccc}
\hline
Loss process & $a \equiv \frac{d\ln t}{d\ln E}$ & $g (p = 2.0)$ & $g (p = 2.2)$ & $g (p = 2.4)$\\
\hline
Synchrotron / Inverse Compton & -1.0 & 1.0     & 1.00018 & 1.0112\\
Bremsstrahlung                & 0.0  & 1.2343  & 1.14299 & 1.07964\\
Ionization                    & 1.0  & 3.00492 & 2.26698 & 1.83161\\
\hline
\end{tabular}
\end{minipage}
\end{table*}

Now we can solve for the proton spectrum given a synchrotron
volumetric luminosity.  Combining equations~\ref{eqn:UpGivenE2Qp},
\ref{eqn:SecondarySpectrum}, and \ref{eqn:nuLnuFromE2Qe}, we have
\begin{equation}
\label{eqn:UpBeforetSynchSubst}
U_p = 12 g^{-1} \left(\frac{E_p}{m_p c^2}\right)^{p - 2} (p - 1) \Upsilon \frac{t_{\pi}}{t_{e} (E_e)} t_{\rm synch} (E_e) f_{\rm sec} \nu \epsilon_{\nu}.
\end{equation}

The pitch-angle averaged synchrotron cooling time is
\begin{equation}
\label{eqn:tSynchPure}
t_{\rm synch} = \frac{6 \pi (m_e c^2)^2}{\sigma_T c E_e B^2},
\end{equation}
where $\sigma_T$ is the Thomson cross section \citep{Rybicki79}.
Substituting it into eqn.~\ref{eqn:UpBeforetSynchSubst}, and using
eqn.~\ref{eqn:nuC} to put $E_e$ in terms of frequency and magnetic
field, we find that
\begin{equation}
\label{eqn:UpAftertSynchSubst}
U_p = \frac{72 \pi m_e c^2}{g \sigma_T c} \left(\frac{20 m_e}{m_p}\right)^{p - 2} \left(\frac{16 m_e c \nu}{3 e}\right)^{(p - 3)/2} B^{-(p+1)/2} (p - 1) \Upsilon \frac{t_{\pi}}{t_{e} (E_e)} f_{\rm sec} \nu \epsilon_{\nu}
\label{eqn:UpFinal}
\end{equation}

\subsection{Which loss process is most important?}
\label{sec:LossTimes} A key ingredient that we need to derive the
equipartition formula is the $e^{\pm}$ cooling time.  Losses from
synchrotron, Inverse Compton (IC), bremsstrahlung, ionization, and
advection all potentially contribute.  The $e^{\pm}$ lifetime from
all of these processes is:
\begin{equation}
t_{e} = [t_{\rm synch}^{-1} + t_{\rm IC}^{-1} + t_{\rm brems}^{-1} + t_{\rm ion}^{-1} + t_{\rm wind}^{-1}]^{-1}.
\end{equation}

Our discussion here is similar to that of \citet{Thompson06},
\citet{Murphy09-zHigh}, and \citet{Lacki10-FRC1}.  We consider M~82
as an example (the other nearby starburst NGC~253 has a
similar environment), which has a magnetic field strength of $\sim
200\ \muGauss$ (from detailed models of the radio spectrum;
\citealt{Domingo05}; \citealt*{Persic08};
\citealt{deCeaDelPozo09-M82,Rephaeli10}), a mean gas density of
$\sim 300\ \cm^{-3}$ (\citealt{Weiss01}, assuming a scale height of
50 pc), and a radiation energy density of $\sim 1000\ \eV\ \cm^{-3}$
(\citealt{Sanders03}, again assuming a starburst radius of 250 pc).
Of course, none of these values are constant throughout the
region, and it is not known how inhomogeneity affects CRs in
starbursts, but using the average values seems to work for models fitting current
radio and gamma-ray data \citep{deCeaDelPozo09-Obs}.  After using
eqn.~\ref{eqn:nuC} to convert between $e^{\pm}$ energy to observed
frequency, the synchrotron lifetime of GHz-emitting $e^{\pm}$ is
\begin{equation}
\label{eqn:tSynch}
t_{\rm synch} = 498\ \kyr\ \left(\frac{B}{200\ \muGauss}\right)^{-3/2} \left(\frac{\nu}{1\ \GHz}\right)^{-1/2},
\end{equation}
from equation~\ref{eqn:tSynchPure}. For a radiation energy density
$U_{\rm rad}$ the IC lifetime is $t_{\rm IC} = 3 \sqrt{3} / (16
\sigma_T U_{\rm rad}) \sqrt{m_e c e B / \nu}$ \citep{Rybicki79}:
\begin{equation}
\label{eqn:tIC}
t_{\rm IC} = 495\ \kyr\ \left(\frac{B}{200\ \muGauss}\right)^{1/2} \left(\frac{\nu}{1\ \GHz}\right)^{-1/2} \left(\frac{U_{\rm rad}}{1000\ \eV\ \cm^{-3}}\right)^{-1}
\end{equation}
The bremsstrahlung lifetime is only very weakly energy dependent,
and at high $e^{\pm}$ energies it is simply
\begin{equation}
\label{eqn:tBrems}
t_{\rm brems} = 104\ \kyr\ \left(\frac{n_H}{300\ \cm^{-3}}\right)^{-1}
\end{equation}
where we assume that $n_H = 10 n_{\rm He}$ in the interstellar
medium \citep{Strong98}. The ionization lifetime is $t_{\rm ion} =
\gamma /  [2.7 c \sigma_T (6.85 + 0.5 \ln \gamma) n_H]$, where
$\gamma = E_e / (m_e c^2) = \sqrt{16 m_e c \nu / (3 e B)}$ is the
$e^{\pm}$ Lorentz factor, and again assuming that $n_H = 10 n_{\rm
He}$ \citep{Schlickeiser02}.  If we evaluate the $\ln\ \gamma$ term
as $\ln 1000$, since 1000 is the approximate Lorentz factor of
GHz-emitting $e^{\pm}$ in starbursts, then we find
\begin{equation}
\label{eqn:tIon}
t_{\rm ion} \approx 178\ \kyr\ \left(\frac{B}{200\ \muGauss}\right)^{-1/2} \left(\frac{\nu}{1\ \GHz}\right)^{1/2} \left(\frac{n}{300\ \cm^{-3}}\right)^{-1}.
\end{equation}
Finally the advection time is roughly the time it takes to cross one
scale height $h$ at the wind speed $v_{\rm wind}$, with wind speeds
of order a few hundred $\kms$ observed in M~82
\citep[e.g.,][]{Greve04,Westmoquette09}.\footnote{We assume
that both the gas and the synchrotron-emitting CRs have the same
scale height $h$, since secondary $e^{\pm}$ are created by
interactions of CR protons with gas and the CR $e^{\pm}$ lifetimes
are short in starbursts.}  Taking $h = 50\ \pc$ and $v_{\rm wind} =
300\ \kms$, we find
\begin{equation}
t_{\rm wind} = 163\ \kyr\ \left(\frac{h}{50\ \pc}\right) \left(\frac{v_{\rm wind}}{300\ \kms}\right)^{-1}.
\end{equation}

For our fiducial values of the physical parameters in M~82, we see
that bremsstrahlung is the most important cooling process for 1~GHz
emitting $e^{\pm}$, though ionization and winds are also important.
However, the final $e^{\pm}$ lifetime is $t_{\rm cool} \approx 39\
\kyr$, shorter than even the bremsstrahlung losses.  We can quantify
the importance of other losses relative to bremsstrahlung with the
ratio
\begin{equation}
X = \frac{t_{\rm brems}}{t_{e}}.
\end{equation}
For our fiducial M~82 values, $X = 2.6$ at 1~GHz; other losses are
even less important.

\begin{figure}
\centerline{\includegraphics[width=8cm]{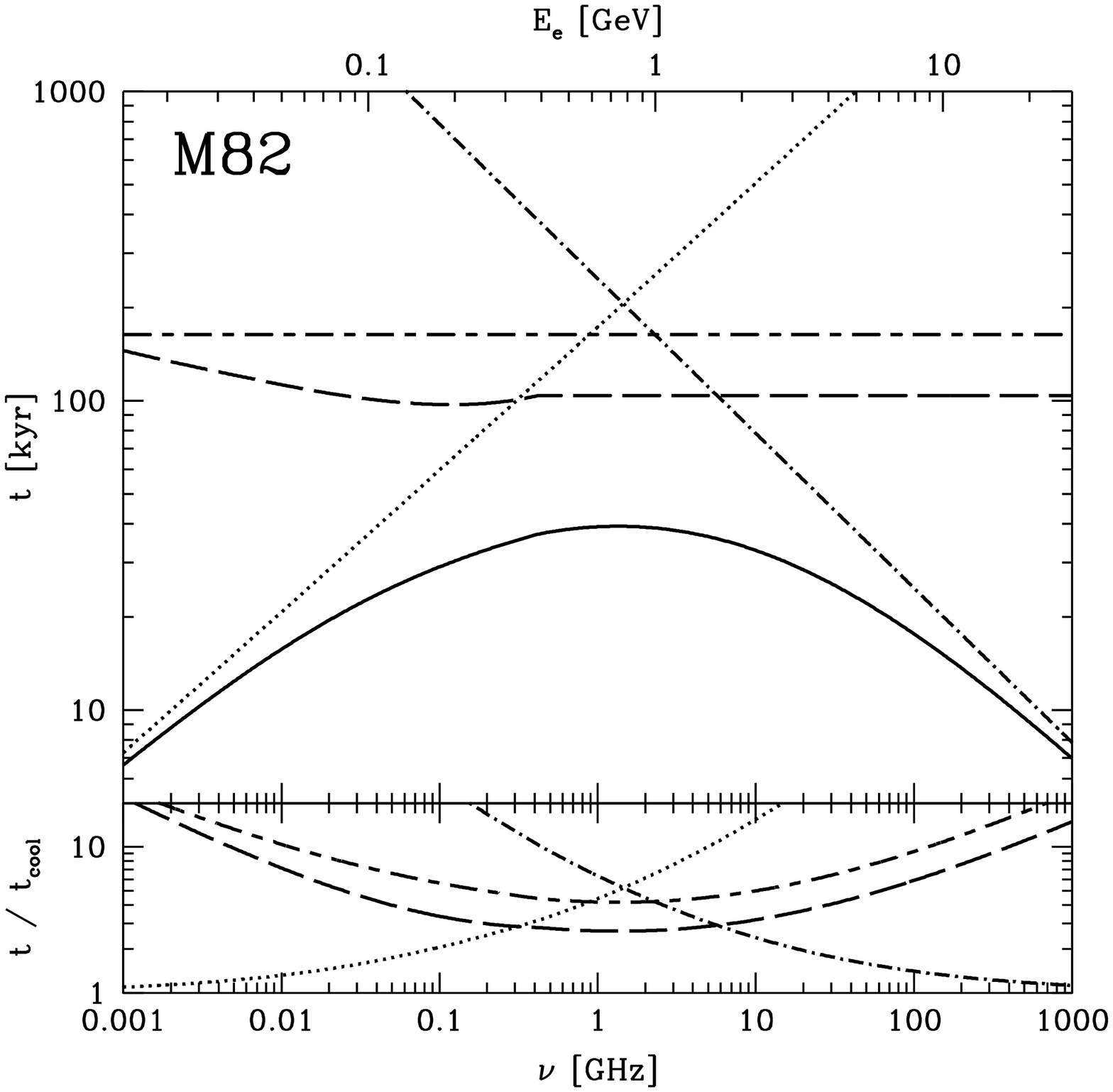}\includegraphics[width=8cm]{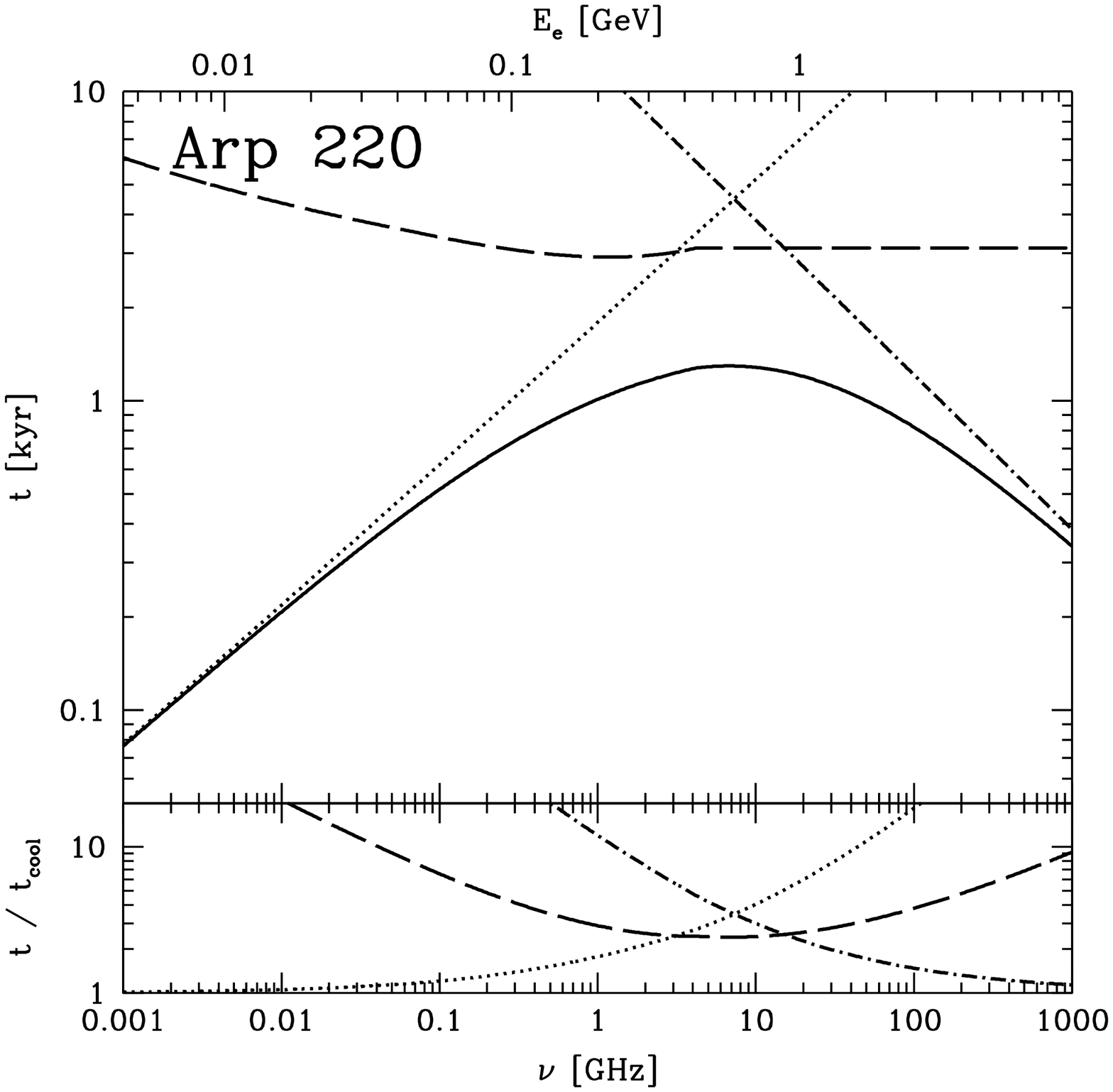}}
\caption{Comparison of the $e^{\pm}$ loss times from different
processes in an M~82-like starburst (left) and in an Arp~220-like
starburst (right).  The total loss time $t_{\rm cool}$ is the solid
line, while bremsstrahlung is long-dashed, ionization is dotted,
wind losses are long/short-dashed, and synchrotron and IC together
are dash-dotted.  On the top are the absolute loss times, while on
the bottom is the ratio of each individual loss time to the total
loss time.\label{fig:M82Losses}\label{fig:A220Losses}}
\end{figure}

A comparison of how the individual loss times compare to each other
at different frequencies is shown in Figure~\ref{fig:M82Losses}
(left).  In this figure, we include the $\ln \gamma$ dependence in
the ionization loss formula, and we use the full bremsstrahlung
losses as given in \citet{Strong98}.\footnote{The $\gamma \la 100$
formula in equation C12 of \citet{Strong98} is erroneously
multiplied by an extra $m_e c^2$, as can be seen in
\citet{Schlickeiser02}, equation 4.4.17.  We have corrected this
when making Figure~\ref{fig:M82Losses}.}  Bremsstrahlung does
dominate the losses of $e^{\pm}$ that emit synchrotron between
300~MHz and 6~GHz, with $X \sim 3$ in this range.  At lower
frequencies, ionization is the most important loss, while at higher
frequencies synchrotron and IC are the most important losses.

We also consider the radio nuclei of Arp~220, using a magnetic field
$\sim 2\ \mGauss$ \citep[e.g.,][]{Lacki10-FRC1}, a density of $\sim
10^4\ \cm^{-3}$ \citep{Downes98}, and a radiation energy density
equivalent to that of a 50~K blackbody ($\sim 30000\ \eV\
\cm^{-3}$). The loss times for these conditions are shown in the
right panel of Figure~\ref{fig:A220Losses}.  At 1~GHz, ionization
now is the most important loss: this is because we see lower energy
$e^{\pm}$ with the higher magnetic field strength.  Near 10~GHz,
though, bremsstrahlung again is the strongest loss, with $X \sim 3$.

For other starbursts, we generally expect advection, bremsstrahlung,
or ionization losses to dominate: advection in low density
starbursts \citep[c.f.,][]{Crocker11-Wild}, ionization in starbursts
with high magnetic fields (likely in the densest starbursts like
Arp~220) where we see low energy $e^{\pm}$, and bremsstrahlung in
intermediate density starbursts.  At frequencies where
bremsstrahlung dominates, $X \approx 3$ (compare with Figure 4, left
panel of \citealt{Lacki10-FRC1}). In starbursts where advection is
very strong, however, protons are likely to escape rather than
produce pionic secondary $e^{\pm}$.  Since advection is equally
strong for protons and $e^{\pm}$, the proton to electron ratio is
just $K_0$ when primaries dominate the $e^{\pm}$ spectrum, and the
revised equipartition formulae of \citet{Beck05} apply.

\subsection{The equipartition and minimum energy formulae}
\label{sec:EquipMinEGiven} For the purposes of this paper, we assume
that the CR energy density entirely consists of protons ($U_{\rm CR}
= U_p$).  Because there are several different energy regimes
governed by different loss processes, the $e^{\pm}$ spectrum is
complex and likely has intrinsic spectral curvature; accounting for
all these loss processes would require knowledge of the density,
radiation energy density, advection speed, diffusion constant, and
magnetic field strength in the volume.  More importantly, because
there are always at least bremsstrahlung losses, and since $t_{\rm
brems} < t_{\pi}$, the steady-state proton/electron ratio will
always be greater than the injection proton/electron ratio (c.f.
equations~\ref{eqn:eSpectrumSoln} and~\ref{eqn:pSpectrumSoln}).  The
injection electron/proton ratio is
\begin{equation}
\frac{Q_e ({\rm prim} + {\rm sec})}{Q_p} < \frac{1}{6} \left(\frac{E_p}{E_{\rm sec}}\right)^{2-p} + \frac{1}{K_0},
\end{equation}
or $\la 18$ per cent for $p = 2.0$, $\la 11$ per cent for $p = 2.2$,
and $\la 7$ per cent for $p = 2.4$.  We therefore conclude that
considering only protons leads to an error of less than $\sim 20$
per cent in $U_p$.  The other uncertainties are generally larger
than this.

Another possible worry of assuming that $U_{\rm CR} = U_p$ is that
the CRs include heavier nuclei, especially helium.  In the Milky
Way, though, the hydrogen/helium ratio at equal rigidities is $\sim
8$, and at equal energies per nucleon, the hydrogen/helium ratio is
$\sim 24$ (\citealt{Webber74}; \citealt*{Webber87}).  This ratio
must be multiplied by the atomic mass $A$ of helium when calculating
the energy density, but still helium is only $(1/24 - 1/8) \times 4
\approx 1/6 - 1/2$ of the CR energy density.  Analysis of the CR
flux at Earth indicate that helium makes up about $\sim 1/6$ of the
Galactic CR energy density, with another $\sim 1/12$ from heavier
nuclei \citep{Webber98}. Furthermore, helium produces pionic
secondary $e^{\pm}$ just as hydrogen does: as far as collisions go,
a helium nucleus at high energies is much like a collection of
protons with the same energy per nucleon.  Nucleons can shield each
other in heavy nuclei, reducing the pp cross section per nucleon as
$A^{-1/4}$ for an atomic mass $A$ \citep{Orth76,Strong98}, but for
helium this just reduces the cross section by 30 per cent.
Therefore, the hadronic starburst radio emission traces high energy
CR helium nuclei as well as CR protons, and our derived $U_p$
basically includes the contribution from helium as well.

The volumetric luminosity $\nu \epsilon_{\nu}$ can be found from
either the total luminosity $\nu L_{\nu}$ or the intensity on a line
of sight $\nu I_{\nu}$.  In the absence of absorption, the
volumetric luminosity is simply:
\begin{equation}
\nu \epsilon_{\nu} = \nu L_{\nu} / V
\end{equation}
where $V$ is the volume of the radio wave emitting region. If the
starburst is inhomogeneous, it is preferable to use the intensity
instead of luminosity to calculate the equipartition field
\citep{Beck05}.  The (unabsorbed) intensity on a line of sight is
given by $dI/d\ell = \epsilon_{\nu} / (4 \pi)$, so that by assuming
a constant $\epsilon_{\nu}$, we can replace the volumetric
luminosity with:
\begin{equation}
\nu \epsilon_{\nu} = 4 \pi \nu I_{\nu} / \ell
\end{equation}
where $\ell$ is the sightline through the radio emitting region.

\subsubsection{Bremsstrahlung-scaled losses}
\label{sec:BremsstrahlungFormula} In general, the lifetime of the CR
$e^{\pm}$ is complex, depending on many loss processes.  It is often
convenient to scale the lifetime to the bremsstrahlung cooling
lifetime.  At GHz frequencies, bremsstrahlung is expected to be an
important or even dominant cooling process for CR $e^{\pm}$ in
starburst galaxies (section~\ref{sec:LossTimes}).  Furthermore, both
pionic and bremsstrahlung losses are mostly independent of energy,
both scale with density, and both are very nearly of the same
magnitude:
\begin{equation}
t_{\pi} \approx 50\ \Myr (n / \cm^{-3})^{-1}
\end{equation}
from \citet{Mannheim94} for the pionic loss time, and (c.f. equation~\ref{eqn:tBrems})
\begin{equation}
t_{\rm brems} \approx 31\ \Myr (n / \cm^{-3})^{-1}
\end{equation}
for the bremsstrahlung lifetime.

\emph{Equipartition formula} -- By setting $B^2 / (8\pi) = \zeta U_p
= \zeta U_{\rm CR}$, the equipartition CR energy density is then
\begin{equation}
\label{eqn:UeqBremsFull}
U_{\rm CR} = \left[\frac{9 m_e c^2}{g \sigma_T c} (8 \pi)^{(3 - p)/4} \left(\frac{20 m_e}{m_p}\right)^{p - 2} \left(\frac{16 m_e c \nu}{3 e}\right)^{(p - 3)/2} (p - 1) \Upsilon \frac{X t_{\pi}}{t_{\rm brems}} f_{\rm sec} \nu \epsilon_{\nu} \zeta^{-(p+1)/4}\right]^{4/(5+p)}
\end{equation}
with a magnetic field of
\begin{equation}
\label{eqn:BeqBremsFull}
B_{\rm eq} = \left[\frac{576 \pi^2 m_e c^2}{g \sigma_T c} \left(\frac{20 m_e}{m_p}\right)^{p - 2} \left(\frac{16 m_e c \nu}{3 e}\right)^{(p - 3)/2} (p - 1) \Upsilon \frac{X t_{\pi}}{t_{\rm brems}} f_{\rm sec} \nu \epsilon_{\nu} \zeta \right]^{2/(5+p)}
\end{equation}
The ratio $\zeta = U_B / U_{\rm CR}$ allows one to consider
departures from exact equipartition.  Here, $X$ is the ratio of
bremsstrahlung cooling time to $t_{\rm cool}$, as previously
defined.  Note that this assumes that $X$ is constant in
energy.  However, $X$ is likely to have some energy dependence, at
least at energies far away from $\sim 1\ \GeV$. Since the frequency
is a function of both energy and magnetic field, the magnetic field
strength would enter implicitly into $X$; in order to solve
self-consistently for $B$, the energy dependence of $X$ must be
included when solving eqn.~\ref{eqn:UpFinal}.  In practice, because
of the weak dependence of $B_{\rm eq}$ on $X$ ($B \propto
X^{2/7.2}$), the effects of energy dependence in $X$ on the
estimated equipartition magnetic field will be small. Furthermore,
below the $\sim\GeV$ energies probed by GHz observations, $dX/dE <
0$ because ionization losses are quicker at low energies, and above
$\sim\GeV$, $dX/dE > 0$ because Inverse Compton and synchrotron
losses are quicker at high energies, so we roughly expect $dX/dE
\approx 0$ (as shown in Figure~\ref{fig:M82Losses}).

Suppose that the starburst is a homogeneous disc, and the
synchrotron-emitting particles fill a fraction $f_{\rm fill}^{\rm
synch}$ of that disk.  If it has radius $R$ and midplane-to-edge
scale height $h$, then the emitting volume is $V = 2 \pi R^2 h
f_{\rm fill}^{\rm CR}$.  The observed flux density $S_{\nu}$ is
$L_{\nu} / (4 \pi D^2)$, where $D$ is the distance.  Scaling to
values typical of nearby starbursts like M~82 and NGC~253:
\begin{eqnarray}
\label{eqn:UeqBrems}
\nonumber U_{\rm CR} & = & 106\ \eV\ \cm^{-3} \left(\frac{\nu}{1\ \GHz}\right)^{\frac{2.4 + 2\Delta p}{7.2 + \Delta p}} \zeta^{-\frac{3.2 + \Delta p}{7.2 + \Delta p}} \times \\
           &   & \left[23.4^{\Delta p} \left(1 + \frac{\Delta p}{1.2}\right) \frac{f_{\rm sec} X}{f_{\rm fill}^{\rm synch}} \left(\frac{S_{\nu}}{\Jy}\right) \left(\frac{R}{250\ \pc}\right)^{-2} \left(\frac{h}{100\ \pc}\right)^{-1} \left(\frac{D}{3.5\ \Mpc}\right)^2  \left(\frac{\Upsilon}{25/6}\right) \left(\frac{g}{1.14}\right)^{-1}\right]^{\frac{4}{7.2 + \Delta p}} \\
\label{eqn:BeqBrems}
\nonumber B_{\rm eq} & = & 65.4\ \muGauss \left(\frac{\nu}{1\ \GHz}\right)^{\frac{1.2 + \Delta p}{7.2 + \Delta p}} \times \\
 & & \left[23.4^{\Delta p} \left(1 + \frac{\Delta p}{1.2}\right) \zeta \frac{f_{\rm sec} X}{f_{\rm fill}^{\rm synch}} \left(\frac{S_{\nu}}{\Jy}\right) \left(\frac{R}{250\ \pc}\right)^{-2} \left(\frac{h}{100\ \pc}\right)^{-1} \left(\frac{D}{3.5\ \Mpc}\right)^2  \left(\frac{\Upsilon}{25/6}\right) \left(\frac{g}{1.14}\right)^{-1}\right]^{\frac{2}{7.2 + \Delta p}}
\end{eqnarray}

\emph{Minimum energy formula} --  The minimum energy magnetic field
is defined as the magnetic field which minimizes $U_B + U_{\rm CR}$
for the observed synchrotron luminosity.  It is found by setting
$d[B^2/(8\pi) + U_p]/dB = 0$.  Numerically, the minimum energy
estimate is very nearly the same as the equipartition estimate:
\begin{eqnarray}
\label{eqn:BMinBremsFull}
B_{\rm min} & = & \left[(p + 1) \frac{144 \pi^2 m_e c^2}{g \sigma_T c} \left(\frac{20 m_e}{m_p}\right)^{p - 2} \left(\frac{16 m_e c \nu}{3 e}\right)^{(p - 3)/2} (p - 1) \Upsilon \frac{X t_{\pi}}{t_{\rm brems}} f_{\rm sec} \nu \epsilon_{\nu}\right]^{2/(p+5)}\\
\nonumber B_{\rm min} & = & 61.3\ \muGauss \left(\frac{\nu}{1\ \GHz}\right)^{\frac{1.2 + \Delta p}{7.2 + \Delta p}} \times \\
& & \left[24.2^{\Delta p} \left(1 + \frac{\Delta p}{1.2}\right)\left(1 + \frac{\Delta p}{3.2}\right) \frac{f_{\rm sec} X}{f_{\rm fill}^{\rm synch}} \left(\frac{S_{\nu}}{\Jy}\right) \left(\frac{R}{250\ \pc}\right)^{-2} \left(\frac{h}{100\ \pc}\right)^{-1} \left(\frac{D}{3.5\ \Mpc}\right)^2  \left(\frac{\Upsilon}{25/6}\right) \left(\frac{g}{1.14}\right)^{-1}\right]^{\frac{2}{7.2 + \Delta p}}
\end{eqnarray}

The ratio $B_{\rm eq} / B_{\rm min}$ is $[4 / (p + 1)]^{2/(p+5)}$, the same as in the revised equipartition formula of \citet{Beck05} (see Table~\ref{table:EqMinRatio}).  When bremsstrahlung losses dominate, the proton to electron ratio is constant with energy, so the proton/electron ratio has the same behavior as if there are no losses.

\begin{table*}
\begin{minipage}{170mm}
\caption{Ratio of equipartition and minimum energy magnetic fields}
\label{table:EqMinRatio}
\begin{tabular}{lccccc}
\hline
Formula & $\displaystyle \frac{B_{\rm eq}}{B_{\rm min}}$ & $\displaystyle \frac{B_{\rm eq}}{B_{\rm min}} (p = 2.0)$ & $\displaystyle \frac{B_{\rm eq}}{B_{\rm min}} (p = 2.2)$ & $\displaystyle \frac{B_{\rm eq}}{B_{\rm min}} (p = 2.4)$ & Reference\\
\hline
Classical formula     & $[4/3]^{2/7}$             & 1.09 & ...  & ...  & \citet{Beck05}\\
Revised formula       & $[4 / (p + 1)]^{2/(p+5)}$ & 1.09 & 1.06 & 1.04 & \citet{Beck05}\\
\hline
\multicolumn{5}{c}{Hadronic formulae}\\
\hline
Bremsstrahlung-scaled losses & $[4 / (p + 1)]^{2/(p+5)}$ & 1.09 & 1.06 & 1.04 & \S~\ref{sec:BremsstrahlungFormula} (Eqns.~\ref{eqn:BeqBremsFull} \&~\ref{eqn:BMinBremsFull})\\
Ionization-dominant losses   & $[4/p]^{2/(p+4)}$         & 1.26 & 1.22 & 1.17 & \S~\ref{sec:IonizationFormula} (Eqns.~\ref{eqn:BeqIonFull} \&~\ref{eqn:BMinIonFull})\\
Synchrotron-dominant losses  & $[4 / (p - 2)]^{2/(p+2)}$ & $\infty ^a$ & 4.2 & 2.8 & \S~\ref{sec:SynchrotronFormula} (Eqns.~\ref{eqn:BeqSynchFull} \&~\ref{eqn:BMinSynch})\\
IC-dominant losses           & $[4 / (p + 2)]^{2/(p+6)}$ & 1.0 & 0.99 & 0.98 & \S~\ref{sec:ICFormula} (Eqns.~\ref{eqn:BEqFullIC} \&~\ref{eqn:BMinFullIC})\\
\hline
\end{tabular}
\\$^a$: As discussed in Section~\ref{sec:SynchrotronFormula}, in practice this
limit can never be attained, because of breaks in the CR spectrum, non
steady-state effects, and other radiative losses that become important when $B$ is low.\\
\end{minipage}
\end{table*}

\subsubsection{Ionization-dominant losses}
\label{sec:IonizationFormula} At low energies, ionization (or
Coulomb losses) will always dominate the energy losses of CR
$e^{\pm}$.  For M~82, ionization losses should become most important
for $e^{\pm}$ radiating below $\sim 300\ \MHz$, while for Arp~220,
ionization losses are most important all the way up to $\sim 3\
\GHz$ (Figure~\ref{fig:M82Losses}).  This assumption is therefore
most appropriate for observations with low frequency telescopes like
LOFAR when they observe starbursts.  While more complicated to scale
from pionic losses, because the ionization losses depend on electron
energy, at sufficiently low energies they will be the only loss,
making the calculation of the electron spectrum relatively simple.

Ignoring logarithmic terms in energy, the ionization loss time for
CR $e^{\pm}$ can be parameterized as:
\begin{equation}
t_{\rm ion} = t_{\rm ion, 0} \left(\frac{B}{B_0}\right)^{-1/2} \left(\frac{\nu}{\nu_0}\right)^{1/2} \left(\frac{n}{\cm^{-3}}\right)^{-1},
\end{equation}
where $t_{\rm ion, 0} = 53\ \Myr$ for $B_0 = 200\ \muGauss$ and
$\nu_0 = 1\ \GHz$ (compare equation~\ref{eqn:tIon}). Likewise, the
pionic loss time can be parametrized as:
\begin{equation}
t_{\pi} = t_{\pi, 0} \left(\frac{n}{\cm^{-3}}\right)^{-1}
\end{equation}
This lets us cancel out the density dependence in the ionization and pionic loss times.

From equation~\ref{eqn:UpFinal}, we have
\begin{equation}
U_p = \frac{72 \pi m_e c^2}{g \sigma_T c} \left(\frac{20 m_e}{m_p}\right)^{p-2} \left(\frac{16 m_e c \nu}{3 e}\right)^{(p-3)/2} B^{-(p+1)/2} (p - 1) \Upsilon \frac{t_{\pi, 0}}{t_{\rm ion, 0}} \left(\frac{B}{B_0}\right)^{1/2} \left(\frac{\nu}{\nu_0}\right)^{-1/2} f_{\rm sec} \nu \epsilon_{\nu}.
\end{equation}

\emph{Equipartition formula} -- Solving for the equipartition
magnetic field $B_{\rm eq} = \sqrt{8 \pi \zeta U_{\rm CR}}$, we find
\begin{equation}
U_{\rm CR} = \left[ \frac{72 \pi m_e c^2}{g \sigma_T c} (8 \pi)^{-p/4} \left(\frac{20 m_e}{m_p}\right)^{p-2} \left(\frac{16 m_e c \nu}{3 e}\right)^{(p-3)/2} (p - 1) \Upsilon \frac{t_{\pi,0}}{t_{\rm ion, 0}} {B_0}^{-1/2} \left(\frac{\nu}{\nu_0}\right)^{-1/2} f_{\rm sec} \nu\epsilon_{\nu} \zeta^{-p/4}\right]^{4/(p+4)}.
\end{equation}
and
\begin{equation}
\label{eqn:BeqIonFull}
B_{\rm eq} = \left[ \frac{576 \pi^2 m_e c^2}{g \sigma_T c} \left(\frac{20 m_e}{m_p}\right)^{p-2} \left(\frac{16 m_e c \nu}{3 e}\right)^{(p-3)/2} (p - 1) \Upsilon \frac{t_{\pi,0}}{t_{\rm ion, 0}} {B_0}^{-1/2} \left(\frac{\nu}{\nu_0}\right)^{-1/2} f_{\rm sec} \nu\epsilon_{\nu} \zeta \right]^{2/(p+4)}.
\end{equation}
For typical starburst values, we have
\begin{eqnarray}
\nonumber U_{\rm CR} & = & 33.6\ \eV\ \cm^{-3} \left(\frac{\nu}{1\ \GHz}\right)^{\frac{0.4 + 2\Delta p}{6.2 + \Delta p}} \zeta^{-\frac{2.2 + \Delta p}{6.2 + \Delta p}} \times \\
           &   & \left[31.2^{\Delta p} \left(1 + \frac{\Delta p}{1.2}\right) \frac{f_{\rm sec} X}{f_{\rm fill}^{\rm synch}} \left(\frac{S_{\nu}}{\Jy}\right) \left(\frac{R}{250\ \pc}\right)^{-2} \left(\frac{h}{100\ \pc}\right)^{-1} \left(\frac{D}{3.5\ \Mpc}\right)^2  \left(\frac{\Upsilon}{25/6}\right) \left(\frac{g}{2.27}\right)^{-1}\right]^{\frac{4}{6.2 + \Delta p}}
\end{eqnarray}
and
\begin{eqnarray}
\nonumber B_{\rm eq} & = & 36.8\ \muGauss \left(\frac{\nu}{1\ \GHz}\right)^{\frac{0.2 + \Delta p}{6.2 + \Delta p}} \times \\
           &   & \left[31.2^{\Delta p} \left(1 + \frac{\Delta p}{1.2}\right) \zeta \frac{f_{\rm sec} X}{f_{\rm fill}^{\rm synch}} \left(\frac{S_{\nu}}{\Jy}\right) \left(\frac{R}{250\ \pc}\right)^{-2} \left(\frac{h}{100\ \pc}\right)^{-1} \left(\frac{D}{3.5\ \Mpc}\right)^2  \left(\frac{\Upsilon}{25/6}\right) \left(\frac{g}{2.27}\right)^{-1}\right]^{\frac{2}{6.2 + \Delta p}}.
\end{eqnarray}

\emph{Minimum energy formula} -- As with the bremsstrahlung loss
formula, we solve for $B_{\rm min}$ by setting $d[B^2/(8\pi) +
U_p]/dB = 0$.  We find
\begin{equation}
\label{eqn:BMinIonFull}
B_{\rm min} = \left[ p \frac{144 \pi^2 m_e c^2}{g \sigma_T c} \left(\frac{20 m_e}{m_p}\right)^{p-2} \left(\frac{16 m_e c \nu}{3 e}\right)^{(p-3)/2} (p - 1) \Upsilon \frac{t_{\pi,0}}{t_{\rm ion, 0}} {B_0}^{-1/2} \left(\frac{\nu}{\nu_0}\right)^{-1/2} f_{\rm sec} \nu \epsilon_{\nu}\right]^{2/(p+4)}.
\end{equation}
Numerically, this comes out to
\begin{eqnarray}
\nonumber B_{\rm min} & = & 30.4\ \muGauss \left(\frac{\nu}{1\ \GHz}\right)^{\frac{0.2 + \Delta p}{6.2 + \Delta p}} \times \\
           &   & \left[34.4^{\Delta p} \left(1 + \frac{\Delta p}{1.2}\right)\left(1 + \frac{\Delta p}{2.2}\right) \frac{f_{\rm sec} X}{f_{\rm fill}^{\rm synch}} \left(\frac{S_{\nu}}{\Jy}\right) \left(\frac{R}{250\ \pc}\right)^{-2} \left(\frac{h}{100\ \pc}\right)^{-1} \left(\frac{D}{3.5\ \Mpc}\right)^2  \left(\frac{\Upsilon}{25/6}\right) \left(\frac{g}{2.27}\right)^{-1}\right]^{\frac{2}{6.2 + \Delta p}}.
\end{eqnarray}

\subsubsection{Synchrotron-dominant losses}
\label{sec:SynchrotronFormula} At high energies, synchrotron and
Inverse Compton losses always dominate the cooling of CR $e^{\pm}$.
For M~82, this should happen for $e^{\pm}$ emitting above $\sim 5\
\GHz$, and for Arp~220, the transition is for $e^{\pm}$ emitting
above $\sim 15\ \GHz$ (Figure~\ref{fig:M82Losses}).  In order for
the observed infrared-radio correlation to hold for starbursts, it
is thought that $U_B \approx U_{\rm rad}$ in starbursts
\citep{Volk89-Calor,Condon91-ULIRGs,Lacki10-FRC1}.  Therefore for
high frequency radio emission, it may be more convenient to scale to
the synchrotron lifetime.  This case was previously considered in
\citet{Pfrommer04}, but we rederive it with our more approximate
approach to compare with the other formulae.  Synchrotron losses do
not depend on density, so the density dependence in $t_{\pi}$ is not
canceled out.

As we can see from equation~\ref{eqn:UpBeforetSynchSubst}, $t_{\rm
cool} = t_{\rm synch}$ cancels the $t_{\rm synch}$ in the CR
$e^{\pm}$ emissivity.  After converting the electron energy $E_e
\approx E_p / 20$ (where a pionic $e^{\pm}$ typically has 1/20 the
energy of its primary proton;
section~\ref{sec:ProtonsToSecondaries}) into synchrotron frequency
using eqn.~\ref{eqn:nuC}, we are left with:
\begin{equation}
\label{eqn:UpForSynch}
U_p = 12 g^{-1} \left(\frac{16 m_e c \nu}{3 e}\right)^{(p-2)/2} B^{-(p-2)/2} \left(\frac{20 m_e}{m_p}\right)^{p-2} (p - 1) \Upsilon t_{\pi} f_{\rm sec} \nu \epsilon_{\nu}
\end{equation}

\emph{Equipartition estimate} -- Setting $B^2/(8\pi) = \zeta U_p = \zeta U_{\rm CR}$, we find:
\begin{equation}
U_{\rm CR} = \left[12 g^{-1} (8\pi)^{-(p-2)/4} \left(\frac{16 m_e c \nu}{3 e}\right)^{(p-2)/2} \left(\frac{20 m_e}{m_p}\right)^{p-2} (p - 1) \Upsilon t_{\pi} f_{\rm sec} \nu \epsilon_{\nu} \zeta^{-(p-2)/4}\right]^{4/(p+2)}.
\end{equation}
The equipartition magnetic field strength is
\begin{equation}
\label{eqn:BeqSynchFull}
B_{\rm eq} = \left[96 \pi g^{-1} \left(\frac{16 m_e c \nu}{3 e}\right)^{(p-2)/2} \left(\frac{20 m_e}{m_p}\right)^{p-2} (p - 1) \Upsilon t_{\pi} f_{\rm sec} \nu \epsilon_{\nu} \zeta\right]^{2/(p+2)}.
\end{equation}

Substituting in the previously used values, we find
\begin{eqnarray}
\nonumber U_{\rm CR} & = & 5.45\ \eV\ \cm^{-3} \left(\frac{\nu}{1\ \GHz}\right)^{\frac{4.4 + 2\Delta p}{4.2 + \Delta p}} \zeta^{-\frac{0.2 + \Delta p}{4.2 + \Delta p}} \times \\
           &   & \left[49.2^{\Delta p} \left(1 + \frac{\Delta p}{1.2}\right) \frac{f_{\rm sec} X}{f_{\rm fill}^{\rm synch}} \left(\frac{S_{\nu}}{\Jy}\right) \left(\frac{R}{250\ \pc}\right)^{-2} \left(\frac{h}{100\ \pc}\right)^{-1} \left(\frac{D}{3.5\ \Mpc}\right)^2 \left(\frac{n_H}{300\ \cm^{-3}}\right)^{-1} \left(\frac{\Upsilon}{25/6}\right) \left(\frac{g}{1.0}\right)^{-1}\right]^{\frac{4}{4.2 + \Delta p}}
\end{eqnarray}
for the energy density and
\begin{eqnarray}
\nonumber B_{\rm eq} & = & 14.8\ \muGauss \left(\frac{\nu}{1\ \GHz}\right)^{\frac{2.2 + \Delta p}{4.2 + \Delta p}} \times \\
           &   & \left[49.2^{\Delta p} \left(1 + \frac{\Delta p}{1.2}\right) \zeta \frac{f_{\rm sec} X}{f_{\rm fill}^{\rm synch}} \left(\frac{S_{\nu}}{\Jy}\right) \left(\frac{R}{250\ \pc}\right)^{-2} \left(\frac{h}{100\ \pc}\right)^{-1} \left(\frac{D}{3.5\ \Mpc}\right)^2 \left(\frac{n_H}{300\ \cm^{-3}}\right)^{-1} \left(\frac{\Upsilon}{25/6}\right) \left(\frac{g}{1.0}\right)^{-1}\right]^{\frac{2}{4.2 + \Delta p}}
\end{eqnarray}
for the equipartition magnetic field strength.

The resulting energy densities and magnetic fields are much smaller
than the bremsstrahlung estimates.  This is because the synchrotron
cooling alone is relatively slow compared to pionic cooling for the
parameters used.  Thus, the pionic $e^{\pm}$ accumulate longer,
leading to a smaller proton/electron ratio.  With less `invisible'
CR proton content compared to CR $e^{\pm}$ content, the energy
density is smaller.

\emph{Minimum energy formula} -- The minimum energy magnetic field strength with these assumptions is
\begin{equation}
\label{eqn:BMinSynch}
B_{\rm min} = \left[24 \pi g^{-1} (p - 2) \left(\frac{16 m_e c \nu}{3 e}\right)^{(p-2)/2} \left(\frac{20 m_e}{m_p}\right)^{p-2} (p - 1) \Upsilon t_{\pi} f_{\rm sec} \nu \epsilon_{\nu}\right]^{2/(p+2)}.
\end{equation}
With our usual parameters, we find that $B_{\rm min}$ is much smaller than $B_{\rm eq}$:
\begin{eqnarray}
\label{eqn:BMinSynchNumer}
\nonumber B_{\rm min} & = & 3.56\ \muGauss \left(\frac{\nu}{1\ \GHz}\right)^{\frac{2.2 + \Delta p}{4.2 + \Delta p}} \left[100^{\Delta p} \left(1 + \frac{\Delta p}{1.2}\right) \left(1 + \frac{\Delta p}{0.2}\right) \frac{f_{\rm sec} X}{f_{\rm fill}^{\rm synch}} \left(\frac{S_{\nu}}{\Jy}\right) \left(\frac{R}{250\ \pc}\right)^{-2} \left(\frac{h}{100\ \pc}\right)^{-1} \right. \\
           &   & \times \left. \left(\frac{D}{3.5\ \Mpc}\right)^2 \left(\frac{n_H}{300\ \cm^{-3}}\right)^{-1} \left(\frac{\Upsilon}{25/6}\right) \left(\frac{g}{1.0}\right)^{-1}\right]^{\frac{2}{4.2 + \Delta p}}
\end{eqnarray}
The reason that $B_{\rm min} \ll B_{\rm eq}$
(Table~\ref{table:EqMinRatio}) is that the derived $U_p$ depends
very weakly on $B$.  When all of the electron power is going into
synchrotron, we have $\nu \epsilon_{\nu} = E_e^2 N_e(E_e) / t_{\rm
synch} = E_e^2 Q_e(E_e) t_{\rm synch} / t_{\rm synch} = E_e^2
Q(E_e)$. Even as $B$ gets arbitrarily small, the assumption that
$e^{\pm}$ cool only by synchrotron implies that, in steady-state,
the synchrotron luminosity is exactly equal to the injected power.
The only way that $B$ affects the synchrotron flux is the dependence
of frequency on magnetic field strength: varying the magnetic field
strength probes different parts of the $e^{\pm}$ spectrum, where
there are different amounts of power injected if $p \ne 2$.  This is
a weak effect.  Therefore, since the derived $U_p$ is nearly
independent of $B$, the nonthermal energy density can usually be
decreased by simply decreasing $U_B$ with little effect on $U_p$.
When $p = 2.0$ exactly, then $B_{\rm min}$ goes to 0.  With a $p =
2$ spectrum, there is equal power injected at all energies, so the
spectral effect disappears.  In practice, this limit is not attained
because (1) there is a high energy cutoff to the CR $e^{\pm}$
spectrum, and the frequency it is observed at depends on $B$; (2)
steady-state may not be attained if synchrotron is the only
radiative loss; and (3) as $B$ decreases, the other losses become
more important; if nothing else, when $B \le 3.5 (1 + z)^2\
\muGauss$, Inverse Compton losses from the CMB exceed synchrotron
losses, so this estimate is no longer valid.  Finally, if
synchrotron really were the only losses, then as $t_{\rm synch} \to
\infty$, $U_e$ becomes bigger than $U_p$, invalidating our
assumption that $U_{\rm CR} \approx U_p$.  In practice, this cannot
happen for hadronic $e^{\pm}$ because bremsstrahlung losses would
always intervene (as discussed in the beginning of the section).

For comparison, \citet{Pfrommer04} found a minimum energy magnetic
field strength for pionic $e^{\pm}$ that can be written as
\begin{equation}
\displaystyle B_{\rm min}^{P04} = \left[\frac{256 \pi^2 (p - 2) (p + 2)}{9 \sqrt{3\pi} (p + 10/3)} \frac{{\cal B} (\frac{p-2}{2}, \frac{3-p}{2}) \Gamma(\frac{p + 8}{4})}{\Gamma(\frac{3p + 2}{12})\Gamma(\frac{3p + 10}{12})\Gamma(\frac{p+6}{4})} \left(\frac{16 m_e}{m_p}\right)^{p-2} \left(\frac{2 \pi m_e c \nu}{3 e}\right)^{(p-2)/2} \frac{\nu \epsilon_{\nu}}{n \sigma_{pp} c}\right]^{2/(p+2)}
\end{equation}
when synchrotron losses dominate and $p \ne 2$, with $\sigma_{pp}$ as the cross section for proton-proton pionic collisions (given in their equation 2.18), ${\cal B}$ is the beta function, and $\Gamma$ is the gamma function.  This version of the estimate assumes there are no primary electrons ($f_{\rm sec} = 1$).  Plugging our fiducial parameters into this equation instead of equation~\ref{eqn:BMinSynch} gives us $3.7\ \muGauss$ instead of $3.6\ \muGauss$ in eqn.~\ref{eqn:BMinSynchNumer}.  The result from our approach is therefore confirmed by \citet{Pfrommer04}.

\subsubsection{IC-dominant losses}
\label{sec:ICFormula} The other dominant loss at high energies is
Inverse Compton (IC). In starburst galaxies, IC losses are expected
to be quick because of intense FIR radiation from star-forming
regions \citep{Condon91-ULIRGs}; the only question is whether the
other losses are stronger (see the discussion in
\ref{sec:LossTimes}; \citealt{Volk89-Calor}).  The IC loss time has
a convenient scaling with the synchrotron loss time:
\begin{equation}
t_{\rm IC} = t_{\rm synch} (U_B / U_{\rm rad})
\end{equation}
where $U_{\rm rad}$ is the radiation energy density.  We therefore have
\begin{equation}
\label{eqn:UpForIC}
U_p = 12 g^{-1} \left(\frac{16 m_e c \nu}{3 e}\right)^{(p-2)/2} B^{-(p-2)/2} \left(\frac{20 m_e}{m_p}\right)^{p-2} (p - 1) \Upsilon t_{\pi} f_{\rm sec} \frac{U_{\rm rad}}{U_B} \nu \epsilon_{\nu}
\end{equation}
These equipartition and minimum energy estimates require both a
density and a radiation field.

\emph{Equipartition estimate} -- By setting the ratio of
magnetic energy density to CR proton energy density as $\zeta$, we
find:
\begin{equation}
U_{\rm eq} = \left[12 g^{-1} (8\pi)^{-(p-2)/4} \left(\frac{16 m_e c \nu}{3 e}\right)^{(p-2)/2} \left(\frac{20 m_e}{m_p}\right)^{p-2} (p - 1) \Upsilon t_{\pi} f_{\rm sec} U_{\rm rad} \nu \epsilon_{\nu} \zeta^{-(p+2)/4}\right]^{4/(p+6)}
\end{equation}
and
\begin{equation}
\label{eqn:BEqFullIC}
B_{\rm eq} = \left[768 \pi^2 g^{-1} \left(\frac{16 m_e c \nu}{3 e}\right)^{(p-2)/2} \left(\frac{20 m_e}{m_p}\right)^{p-2} (p - 1) \Upsilon t_{\pi} f_{\rm sec} U_{\rm rad} \nu \epsilon_{\nu} \zeta \right]^{2/(p+6)},
\end{equation}
with the accompanying scalings of:
\begin{eqnarray}
\nonumber U_{\rm CR} & = & 69.3\ \eV\ \cm^{-3} \left(\frac{\nu}{1\ \GHz}\right)^{\frac{4.4 + 2\Delta p}{8.2 + \Delta p}} \zeta^{-\frac{4.2 + \Delta p}{8.2 + \Delta p}} \left[26.1^{\Delta p} \left(1 + \frac{\Delta p}{1.2}\right) \frac{f_{\rm sec} X}{f_{\rm fill}^{\rm synch}} \left(\frac{S_{\nu}}{\Jy}\right) \left(\frac{R}{250\ \pc}\right)^{-2} \left(\frac{h}{100\ \pc}\right)^{-1} \right.\\
           &   & \times \left. \left(\frac{D}{3.5\ \Mpc}\right)^2 \left(\frac{n_H}{300\ \cm^{-3}}\right)^{-1} \left(\frac{U_{\rm rad}}{1000\ \eV\ \cm^{-3}}\right) \left(\frac{\Upsilon}{25/6}\right) \left(\frac{g}{1.0}\right)^{-1}\right]^{\frac{4}{8.2 + \Delta p}}
\end{eqnarray}
and
\begin{eqnarray}
\nonumber B_{\rm eq} & = & 52.8\ \muGauss \left(\frac{\nu}{1\ \GHz}\right)^{\frac{2.2 + \Delta p}{8.2 + \Delta p}} \left[26.1^{\Delta p} \left(1 + \frac{\Delta p}{1.2}\right) \zeta \frac{f_{\rm sec} X}{f_{\rm fill}^{\rm synch}} \left(\frac{S_{\nu}}{\Jy}\right) \left(\frac{R}{250\ \pc}\right)^{-2} \left(\frac{h}{100\ \pc}\right)^{-1} \right.\\
           &   & \times \left. \left(\frac{D}{3.5\ \Mpc}\right)^2 \left(\frac{n_H}{300\ \cm^{-3}}\right)^{-1} \left(\frac{U_{\rm rad}}{1000\ \eV\ \cm^{-3}}\right) \left(\frac{\Upsilon}{25/6}\right) \left(\frac{g}{1.0}\right)^{-1}\right]^{\frac{2}{8.2 + \Delta p}}.
\end{eqnarray}

\emph{Minimum energy estimate} -- Using eqn.~\ref{eqn:UpForIC} to
find the minimum nonthermal energy density, we calculate
\begin{equation}
\label{eqn:BMinFullIC}
B_{\rm min} = \left[(p+2) 192 \pi^2 \left(\frac{16 m_e c \nu}{3 e}\right)^{(p-2)/2} \left(\frac{20 m_e}{m_p}\right)^{p-2} (p - 1) \Upsilon t_{\pi} f_{\rm sec} U_{\rm rad} \nu \epsilon_{\nu}\right]^{2/(p+6)},
\end{equation}
or
\begin{eqnarray}
\label{eqn:BMinICNumer}
\nonumber B_{\rm min} & = & 53.5\ \muGauss \left(\frac{\nu}{1\ \GHz}\right)^{\frac{2.2 + \Delta p}{8.2 + \Delta p}} \left[25.9^{\Delta p} \left(1 + \frac{\Delta p}{1.2}\right) \left(1 + \frac{\Delta p}{4.2}\right) \frac{f_{\rm sec} X}{f_{\rm fill}^{\rm synch}} \left(\frac{S_{\nu}}{\Jy}\right) \left(\frac{R}{250\ \pc}\right)^{-2} \left(\frac{h}{100\ \pc}\right)^{-1} \right.\\
           &   & \times \left. \left(\frac{D}{3.5\ \Mpc}\right)^2 \left(\frac{n_H}{300\ \cm^{-3}}\right)^{-1} \left(\frac{U_{\rm rad}}{1000\ \eV\ \cm^{-3}}\right) \left(\frac{\Upsilon}{25/6}\right) \left(\frac{g}{1.0}\right)^{-1}\right]^{\frac{2}{8.2 + \Delta p}}.
\end{eqnarray}
for $p = 2.2 + \Delta p$.  For $p$ near 2, the minimum energy and
equipartition magnetic field strengths are nearly identical in this
case (Table~\ref{table:EqMinRatio}).

\citet{Pfrommer04} also derived the minimum energy estimate for
IC-loss dominated pionic $e^{\pm}$ in a more sophisticated manner,
finding:
\begin{equation}
\displaystyle B_{\rm min}^{P04} = \left[\frac{2048 \pi^3 (p + 2)^2}{9 \sqrt{3\pi} (p + 10/3)} \frac{{\cal B} (\frac{p-2}{2}, \frac{3-p}{2}) \Gamma(\frac{p + 8}{4})}{\Gamma(\frac{3p + 2}{12})\Gamma(\frac{3p + 10}{12})\Gamma(\frac{p+6}{4})} \left(\frac{16 m_e}{m_p}\right)^{p-2} \left(\frac{2 \pi m_e c \nu}{3 e}\right)^{(p-2)/2} \frac{\nu \epsilon_{\nu} U_{\rm rad}}{n \sigma_{pp} c}\right]^{2/(p+6)}.
\end{equation}
Here, we have substituted the general radiation energy density
$U_{\rm rad}$ for the CMB energy density.  Plugging our fiducial
starburst parameters into this equation gives us $55\ \muGauss$ as
the coefficient for equation~\ref{eqn:BMinICNumer}.  Like the
synchrotron-loss dominant formula, the results of \citet{Pfrommer04}
are in accord with our own.

\section{Applying the Equipartition Formula}
\subsection{How much of the radio flux is diffuse and nonthermal?}
So far, we have been assuming throughout the paper that all of
a starburst's GHz radio flux is diffuse synchrotron emission from CR
$e^{\pm}$ in the interstellar medium.  We also assumed that the
synchrotron flux is transmitted freely to Earth.  Neither is
completely true, although it turns out these are often reasonable
approximations.

At high frequencies, thermal free-free emission, which does not
trace magnetic fields or CRs, becomes increasingly important, since
it has a flat $\nu^{-0.1}$ spectrum.  However, the total amount of
free-free emission is limited by the number of ionizing photons
generated by the starburst.  At 1.4~GHz, the thermal fraction is at
most $\sim 1/8$ for starbursts that lie on the observed
infrared-radio correlation, as estimated by \citet{Condon92}; if
many ionizing photons escape the starburst or are destroyed by dust
do not contribute, the thermal fraction might be smaller.  A few
very young starbursts appear to have no synchrotron flux, perhaps
because supernovae have not gone off in them \citep{Roussel03}, but
most star-forming galaxies and starbursts have radio spectral
indices $\ll -0.1$, indicating nonthermal emission dominates the GHz
flux
\citep[e.g.,][]{Klein88,Condon92,Niklas97-FF,Clemens08,Ibar10,Williams10}.
Radio spectrum fits to normal spiral galaxies indicate their typical
1.4~GHz thermal fraction is $\sim 10\%$ \citep*{Niklas97-FF}.  For
starburst galaxies, the spectral fits are more difficult, because
the transition between the different cooling processes at different
frequencies cause the intrinsic synchrotron spectrum to steepen
\citep{Thompson06}, whereas free-free emission flattens the
spectrum; disentangling the effects of the two is difficult
\citep{Condon92}.  Model fits to starburst radio spectra so far generally indicate
GHz thermal fractions of a few percent (e.g., \citealt*{Klein88};
\citealt{Williams10}).  We adopt $f_{\rm therm} = [9 (\nu /
\GHz)^{-0.6} + 1]^{-1}$, which is appropriate if the nonthermal
synchrotron spectrum has a $\nu^{-0.7}$ spectrum and the thermal
fraction at 1~GHz is 10\%.

The same ionized matter that generates free-free emission must also
have free-free absorption on some level, but the effects of
absorption are harder to calculate because it depends on the
geometry of the ionized gas.  The simplest approach to estimate the
free-free absorption from the radio spectrum is with uniform slab
models, in which the synchrotron-emitting CRs and ionized $10^4$ K
gas are assumed to be homogeneous and fully mixed.  With uniform
slab models, the free-free turnovers are generally expected to be in
the range of a few hundred MHz for M~82 and NGC~253's starburst
cores, with optical depths of order $\sim 0.1$ at 1 GHz 
\citep{Carilli96,Williams10,Adebahr12}.  Free-free absorption may be
important for frequencies up to a few GHz in Ultraluminous Infrared
Galaxies
\citep[ULIRGs,][]{Condon91-ULIRGs,Torres04-Arp220,Clemens08}.  In
either case, free-free absorption might significantly suppress the
observed synchrotron luminosity at frequencies where ionization
cooling dominates

Yet the ionized gas is not likely to be uniformly distributed in the
starburst.  Much of the ionized mass instead is likely concentrated
discrete H\,II regions, which are dense but have small filling factor.
Since not all sightlines through the starburst necessarily pass
through an H\,II region, it is possible some synchrotron flux is
transmitted even as the frequency descends deep into the MHz range
\citep{Lacki12-LowNu}.  Radio recombination line observations are
often interpreted with models of H\,II regions.  These studies find H
II regions of a wide range of densities in starbursts, but generally
fall into the categories of relatively low density ($\sim 1000\
\cm^{-3}$) regions with relatively high covering fractions and
relatively high density ($\sim 10^5\ \cm^{-3}$) regions with
relatively low covering fractions
\citep{Anantharamaiah93,RodriguezRico04,RodriguezRico06}.  In
Arp~220, the high density H\,II regions are opaque even past 10~GHz,
but they cover only a fraction of a percent of the starburst; a more
diffuse component of ionized gas has an optical depth of $\sim 10\%$
at 8~GHz \citep{Anantharamaiah00}.

Given the difficulty of accurate calculations of free-free
absorption, the relatively weak dependence of the equipartition and
minimum energy estimates on the flux, the uncertainty in other
factors like emitting volume (both because of projection effects and
because of the unknown value of the CR filling factor), and the fact
that free-free absorption optical depth is expected to be
significantly less than 1 at the frequencies we consider (1 -
1.4~GHz for most starbursts, 8.4~GHz for ULIRGs), we ignore it when
estimating the magnetic fields and CR energy densities below.
However, its effects should be considered carefully when working at
low frequencies with the ionization-dominant losses formulae.

Finally, not all of the synchrotron flux comes from the CRs in the
starburst interstellar medium.  First, some of the radio emission
actually comes from individual radio sources like supernovae
remnants.  The fraction of the flux from these individual sources is
only a few percent of the total starburst flux, though, so we ignore
it \citep{Lisenfeld00,Lonsdale06}.  Second, not all of the flux
comes from the starbursting region itself, where the CRs are
accelerated or created through pion productions, and where the
magnetic and CR energy densities are though to be highest.  Some
comes from the surrounding host galaxy, and additional emission can
arise from a ``halo'' region thought to arise when CR $e^{\pm}$ are
advected away from the galaxy.  In the case of NGC~253, about half
of the radio flux in fact comes from these regions rather than the
starburst core, so we use the core flux only
\citep{Heesen09,Williams10}.  Likewise, in NGC~1068, most of the
radio flux comes from an active nucleus and its jets, so we use the
estimated flux of the starburst alone \citep{WynnWilliams85}.  We
assume in the other cases that all of the synchrotron flux comes
from the starburst core, but this can be tested with resolved
observations.  This is known to be a reasonable approximation for
M~82, where the starburst core does actually dominate the total flux
at frequencies above 1~GHz \citep{Adebahr12}.

\subsection{New equipartition estimates of $B$ in starbursts}
We present new equipartition estimates of the magnetic field
strength in selected starburst galaxies in
Table~\ref{table:NewEquip}.  These estimates use the
bremsstrahlung-dominated formula with $X = 3$, and also assume that
$p = 2.2$ and $f_{\rm fill}^{\rm CR} = f_{\rm sec} = \zeta = 1$. For
comparison, we also give the results for the classical and revised
equipartition formula.  Note that we use 8.4~GHz radio data for the
ULIRGs Arp~220, Arp~193, and Mrk~273; at these higher frequencies,
bremsstrahlung still should be the most important loss
(Figure~\ref{fig:A220Losses}, right panel).

\begin{table*}
\begin{minipage}{170mm}
\caption{Updated equipartition estimates for starbursts}
\label{table:NewEquip}
\begin{tabular}{lcccccccccccc}
\hline
Starburst & $D$ & $R$ & $h$ & $\nu$ & $S_{\nu}$ & \multicolumn{3}{c}{Bremsstrahlung-scaled$^a$} & Classical$^b$ & \multicolumn{2}{c}{Revised$^c$} & References$^d$\\
          &     &     &     &       &          & $U_{\rm CR}$ & $B_{\rm eq}$ & $B_{\rm min}$ & $B_{\rm min}$ & $B_{\rm eq}$ & $B_{\rm min}$ & \\
          & (Mpc) & (pc)  & (pc)  & (GHz) & (Jy) & (eV cm$^{-3}$) & ($\muGauss$) & ($\muGauss$) & ($\muGauss$) & ($\muGauss$) & ($\muGauss$)\\
\hline
M 82         & 3.6  & 250 & 50 & 1.0 & 8.94      & 940  & 190 & 180 & 160 & 240 & 220 & (1)\\
NGC 253 Core & 3.5  & 150 & 50 & 1.0 & 3.0       & 880  & 190 & 180 & 160 & 230 & 220 & (1)\\
NGC 4945     & 3.7  & 540 & 50 & 1.4 & 4.2       & 300  & 110 & 100 & 89  & 130 & 130 & (2)\\
NGC 1068 Starburst & 13.7 & 3000 & 50 & 1.4  & 1 & 86   & 59  & 55  & 47  & 72  & 68  & (3)\\
IC 342       & 4.4  & 710 & 50 & 1.4 & 2.25      & 190  & 87  & 82  & 71  & 110 & 100 & (4)\\
NGC 2146     & 12.6 & 520 & 50 & 1.4 & 1.09      & 570  & 150 & 140 & 130 & 190 & 180 & (4)\\
NGC 3690     & 42.2 & 2460 & 50 & 1.4 & 0.66     & 300  & 110 & 100 & 89  & 130 & 130 & (4)\\
NGC 1808     & 14.2 & 1030 & 50 & 1.4 & 0.52     & 200  & 91  & 85  & 73  & 110 & 100 & (4)\\
NGC 3079     & 15.9 & 190 & 50 & 1.4 & 0.85      & 2000 & 280 & 270 & 240 & 350 & 330 & (4)\\
Arp 220 West & 75   & 100 & 50 & 8.4 & 0.072     & 9300 & 610 & 580 & 500 & 750 & 710 & (5)\\
Arp 220 East & 75   & 100 & 50 & 8.4 & 0.061     & 8500 & 590 & 550 & 480 & 720 & 670 & (5)\\
Arp 193      & 99   & 160 & 50 & 8.4 & 0.035     & 5000 & 450 & 420 & 360 & 550 & 520 & (4)\\
Mrk 273      & 160  & 160 & 50 & 8.4 & 0.044     & 9800 & 630 & 590 & 510 & 770 & 720 & (4)\\
\hline
\end{tabular}
\\
Columns: $D$ is the distance to the starburst; $R$ is the assumed radius of the starburst disc and $h$ is the assumed midplane-to-edge scale height of the starburst disc, for a total volume of $V = 2 \pi R^2 h$; $S_{\nu}$ is the total observed flux density at frequency $\nu$.  We take the thermal fraction of the radio emission to be $f_{\rm therm} = [9 (\nu / \GHz)^{-0.6} + 1]^{-1}$.\\
$^a$: Equipartition energy densities, equipartition magnetic field strengths, and minimum energy magnetic field strengths calculated assuming radio emission is hadronic and with $e^{\pm}$ lifetimes scaled to the bremsstrahlung loss time, from equations~\ref{eqn:UeqBremsFull}, \ref{eqn:BeqBrems}, and~\ref{eqn:BMinBremsFull}.  We assume $X = 3$, $f_{\rm sec} = 1$, $f_{\rm fill}^{\rm CR} = 1$, $\zeta = 1$, and $p = 2.2$.  \\
$^b$: Classical minimum energy estimate of magnetic field strength from \citet{Longair10}, which assumes $K_0 = 100$, $\nu_{\rm min} = \nu$, and $\alpha = 0.75$.\\
$^c$: Revised equipartition and minimum energy magnetic field strengths from \citet{Beck05}, which assumes $p = 2.2$ (synchrotron spectral index $\alpha = 0.6$) and $K_0 = 100$.\\
$^d$: References -- (1): M~82 and NGC~253 radio flux densities from \citet{Williams10}; (2): NGC~4945 radio flux density and size from \citet{Strickland04}; (3): Radio flux density and approximate radius of NGC~1068 starburst from \citet*{WynnWilliams85}; (4): Radio flux densities and radii compiled from \citet{Thompson06}; (5): Distance and sizes of Arp~220's radio nuclei from \citet{Sakamoto08}; radio flux density from \citet{Thompson06}.\\
\end{minipage}
\end{table*}

The equipartition magnetic field strengths range from 60--600~$\
\muGauss$.  For M~82 and NGC~253, the estimated magnetic fields are
$200\ \muGauss$, which is entirely in line with broadband modelling
of their nonthermal spectra \citep[e.g.,][]{deCeaDelPozo09-Obs}.
\citet{Klein88} and \citet{Adebahr12} found smaller magnetic
field strengths of 50--100$\ \muGauss$, but they assumed the
emission was spread over a larger volume so that the energy
densities were smaller.  The magnetic field strength estimates for
the distant ULIRGs such as Arp~220 are of order half a milliGauss,
which is several times smaller than the few milliGauss expected from
the linearity of the FIR-radio correlation
\citep{Condon91-ULIRGs,Lacki10-FRC1} and detailed modeling of
Arp~220's CR population \citep{Torres04-Arp220}.  The derived
magnetic field strengths depend weakly on $\zeta = U_B / U_{\rm
CR}$, though, with $B \propto \zeta^{2/(5 + p)}$ and $U_{\rm CR}
\propto \zeta^{-(1 + p)/(5 + p)}$ for the bremsstrahlung-loss
formulae.  At a given radio flux, and if no other assumptions are
changed, magnetic field strengths that are 3 times stronger than
equipartition require $\zeta \approx 50$ and cosmic ray energy
densities  6 times smaller than the values derived for $\zeta = 1$.

However, the most interesting feature of the new equipartition
estimates evident in Table~\ref{table:NewEquip} is that they are
still of the same order as the classical and revised equipartition
and minimum-energy estimates.

\subsection{Why do previous equipartition formulae seem to work?}
If previous equipartition formulae do not apply to starbursts,
because of starbursts' strong electron cooling and the presence of
secondary $e^{\pm}$, why do these formulae give similar results to
the corrected formula?

The answer is that the steady-state proton/electron ratio at $\sim
\GeV$ energies is not far from the canonical $\sim 100$ in
starbursts, although the reasons for this factor are more subtle
than usually realized.  The steady-state proton/electron number
ratio at any given energy is
\begin{equation}
K = \frac{N_p (E)}{N_e (E)} = \frac{6 t_{\pi} X (p - 1) f_{\rm sec}}{F_{\rm cal} t_{\rm brems}} \left(\frac{E_p}{E_{\rm sec}}\right)^{p - 2}
\end{equation}
In a starburst like M~82 with $F_{\rm cal} = 0.4$, $p = 2.2$, and $X
= 3$ (and therefore $f_{\rm sec} = 0.73$), we find $K = 120$.  Even
for $F_{\rm cal} = 1$, $K = 57$.  Therefore, an estimate of the
proton energy density that assumes a constant proton/electron ratio
of $K = 100$ is actually correct to order of magnitude for 1~GHz
observations.

The reasons the assumption works are that:
\begin{itemize}
\item While synchrotron and IC strongly suppress the proton/electron
ratio at high energy, and while ionization losses strongly suppress
the proton/electron ratio at low energy, bremsstrahlung-dominated
losses roughly preserve the proton/electron ratio, since $t_{\pi} /
t_{\rm brems} \approx 1.6$.  Although bremsstrahlung is more
important at high density, reducing the steady-state electron
density \citep{Beck05}, the pionic losses also scale with density,
reducing the steady-state proton density.

\item At 1~GHz, we are observing at intermediate energies in
starbursts, a sweet spot where none of the other losses greatly
overpowers bremsstrahlung.  But while bremsstrahlung may be the
strongest individual loss, the other losses combined are more
important than bremsstrahlung ($X \approx 3$).  As seen in
Figure~\ref{fig:M82Losses}, the $e^{\pm}$ radiative lifetime reaches
a maximum for energies observed at GHz frequencies.  So the combined
effect of the losses is to suppress the proton/electron ratio, but
only by a factor of $\sim 5$.

\item On the other hand, the proton/electron injection ratio is
raised by a factor $\sim 5$ due to the presence of pionic
secondaries. The `dilution' effect arising because secondary
$e^{\pm}$ are injected at much lower energy than the primary protons
helps ensure that the secondaries do not swamp the electron
population.
\end{itemize}
This is related to the `high-$\Sigma_g$ conspiracy' posited by
\citealt{Lacki10-FRC1} to explain the FIR-radio correlation from
normal galaxies to starbursts: radio emission is enhanced by
secondaries, but is suppressed by non-synchrotron losses including
bremsstrahlung.

A consequence is that if we move to higher or lower electron energy
(or observing frequency), $X$ will increase beyond $\sim 3$.  The
coincidence no longer works, and the proton/electron ratio is
increased by either ionization losses at low frequency or
synchrotron and IC losses at high frequency.  At sufficiently low
energies ($\la 100\ \MeV$; observed $\la 40\ \MHz$ in the M~82 and
NGC~253 starbursts), the pionic secondary spectrum will also subside
due to the kinetics of pion production.  At low frequencies, such as
those observed by LOFAR, both of these effects can increase the
proton/electron ratio above that of the hadronic
bremsstrahlung-losses case.  At a few hundred MHz, where the
secondaries are present, the ionization-loss formulae
(section~\ref{sec:IonizationFormula}) should work since it already
takes into account the changing proton/electron ratio.  The
assumption of a CR proton injection spectrum of the form of a $E^{-p}$
power law also probably starts to break down at proton energies of a
GeV as well, with a momentum power law or some non-linear
acceleration spectrum likely more appropriate.

However, while these effects formally invalidate the
bremsstrahlung-loss dominated formula outside of the frequency range
where bremsstrahlung losses dominate (300~MHz to 6~GHz for M~82;
3~GHz to 15~GHz in Arp~220's nuclei), in practice, their effects are
relatively benign.  From the classical and revised formulae, the
proton/electron ratio affects the magnetic field strength estimates
only approximately as $K_0^{2/7.2}$: an order of magnitude change in
$K_0$ only results in a factor $\sim 2$ change in $B_{\rm min}$ or
$B_{\rm eq}$.  At low frequencies, the ratio of ionization to
bremsstrahlung loss times only goes as $t_{\rm ion}/t_{\rm brems}
\propto \nu^{1/2}$, and the proton/electron ratio varies similarly.
To take an extreme example, M~82's radio flux at 22.5~MHz is 39~Jy
\citep*{Roger86}.  Naively putting these values into the
bremsstrahlung loss formula (and assuming all of this flux comes
from the starburst region) gives $B_{\rm eq} = 170\ \muGauss$ and
$B_{\rm min} = 160\ \muGauss$ -- values that are within $\sim 25$
per cent of those obtained using the formula at 1~GHz.  The more
formally correct ionization-loss formulae instead give $B_{\rm eq} =
180\ \muGauss$ and $B_{\rm min} = 140\ \muGauss$.

\section{Conclusion}
We rederive minimum energy and equipartition estimates for the
magnetic fields for the conditions that prevail in starburst
galaxies.  In these galaxies, the radio-emitting CR $e^{\pm}$
population may consist largely of hadronic $pp$ secondaries rather
than primaries.  Furthermore, strong radiative losses, particularly
bremsstrahlung at GeV energies, set the lifetime of the CR $e^{\pm}$
and affect the proton/electron ratio.  Despite these effects, we
found that the steady-state proton/electron ratio is probably close
to $\sim 100$ for GHz-emitting $e^{\pm}$, assuming that $t_{\rm
brems} \approx 3 t_{e}$.  As a result, the classical and revised
equipartition formulae give results that are quite similar to ours,
despite their different assumptions.

Although we scale the CR $e^{\pm}$ lifetime to individual losses
(particularly bremsstrahlung) to derive an analytical result, it
should also be possible to numerically solve for $B_{\rm eq}$ and
$B_{\rm min}$ using the full expression for $t_{\rm cool}$ including
all losses, if equation~\ref{eqn:UpAftertSynchSubst} is used.
However, since the density would not fully cancel out (due to
synchrotron and IC losses in the $e^{\pm}$ lifetime), and because
the IC lifetime depends on the radiation energy density, such an
estimate would require both the gas density the CRs experience and
the radiation energy density.

While our focus has been on starburst galaxies, the formulae above
should apply for any radio source where the CR $e^{\pm}$ population
is mostly pionic $pp$ secondaries and is strongly cooled.
\citet{Pfrommer04} already considered the case of radio emission
from galaxy clusters, which may come from secondary $e^{\pm}$
\citep{Dennison80} that cool from synchrotron and IC losses off the
CMB over the Gyr lifetimes of clusters.  Active Galactic Nuclei may
also accelerate CR protons when their jets are stalled by
surrounding gas \citep[e.g.,][]{AlvarezMuniz04}.  These protons may
interact with the gas to produce pionic secondaries, which may
radiate in the radio; the formulae derived here would also apply in
such a case.

\section*{Acknowledgments}
BCL was supported by a Jansky fellowship from the National Radio
Astronomy Observatory (NRAO).  NRAO is operated by Associated
Universities, Inc., under cooperative agreement with the National
Science Foundation. RB acknowledges financial support from grant DFG
FOR1254. BCL acknowledges earlier discussions with Todd
Thompson.  We also thank Marita Krause for comments on the
manuscript.

\end{document}